\documentclass[aps, reprint]{revtex4-1}

\usepackage[utf8]{inputenc}
\usepackage{t1enc}
\usepackage{graphicx}   % need for figures
\usepackage{amsmath}

\begin{document}

\title{Broadband magnetoelastic coupling in magphonic crystals for high-frequency nanoscale spin wave generation}
\author{Piotr Graczyk}
\email{graczyk@amu.edu.pl}
\author{Jarosław Kłos}
\author{Maciej Krawczyk}
\email{krawczyk@amu.edu.pl}
\affiliation{Faculty of Physics, Adam Mickiewicz University in Poznan, Umultowska 85, 61-614 Poznan, Poland}

\begin{abstract}

Spin waves are promising candidates for information carriers in advanced technology. 
The interactions between spin waves and acoustic waves in magnetic nanostructures 
are of much interest because of their potential application for 
spin wave generation, amplification and transduction.
We investigate numerically the dynamics of magnetoelastic excitations in a one-dimensional magphonic crystal 
consisting of alternating layers of permalloy and cobalt.
We use the plane wave method and the finite element method for frequency- and time-domain simulations, respectively. 
The studied structure is optimized for hybridization of 
specific spin-wave and acoustic dispersion branches 
in the entire Brillouin zone in a broad frequency range. 
We show that this type of periodic structure can be used for efficient generation of high-frequency spin waves.

\end{abstract}

\maketitle

Magnonics, alongside spintronics, opens the possibility to design devices 
even smaller, faster and more energetically efficient than electronic ones \cite{Nikinov2013,Amaru2013,Kos2016}. 
This concerns signal generation, transmission, amplification and manipulation as well as data storage \cite{Demidov2016,Chumak2015,Dutta2015,Gertz2016}. 
Also the numerous possibilities of tuning the properties of spin waves makes them promising candidates for information carriers. 
A way to steer spin waves is to take advantage of the reprogrammability of magnetic materials  \cite{Krawczyk2014}.

An approach in reprogrammable magnonics is based on strain and magnetoelastic interaction \cite{Filimonov2002}. 
Static strain can tune the frequency of spin waves~\cite{Graczyk2016} 
and affect the dispersion and band gap widths of magnonic and phononic crystals 
\cite{BouMatar2012, Ding2014, Robillard2009, Vasseur2011, Zhang2012}. 
On the other hand, dynamic strain, i.e. acoustic waves, 
can have a significant influence on the dynamics of magnetic moments 
and even lead to magnetization reversal~\cite{Sampath2016, Chudnovsky2016}. 
Magnetoelastic interaction between phonons and magnons can also open a new way of spin wave generation~\cite{Gowtham2016}, 
hence the vital importance of the research on the enhancement of this interaction.
It seems especially attractive to study periodic composites, 
which can provide a basis for the application of magnetoelastic coupling in high-frequency devices.
A periodic medium which combines the characteristics of phononic and magnonic crystals is referred to as a magphonic crystal \cite{Pan2013}.

From the perspective of applications it is desirable to operate on
high-frequency, short-wavelength spin waves 
to obtain ultrafast, broad bandwidth, highly miniaturized magnonic devices. 
Planar spin waves or spin-wave beams can be generated by electromagnetic wave transducers~\cite{Vlaminck2010, Demidov2009,  Gruszecki2015, Pirro2014}. 
However, as the wavelength of spin waves is related to the dimensions of the antenna, 
this method has limitations, which can be overcome 
by combining the antenna with a periodic array of magnetic nanodisks~\cite{Yu2016}. 
A number of other approaches to spin wave generation have been proposed recently, 
based on spin-polarized electric currents~\cite{Bonetti2009, Madami2011, Madami2015}, 
acoustic wave-induced spin currents~\cite{Uchida2011}, 
pure spin currents~\cite{Demidov2016}, alternating voltage~\cite{Cherepov2014}, 
oscillating domain walls~\cite{VandeWiele2016}, magnetic vortex cores~\cite{Wintz2016} 
or demagnetizing fields at nanostructure edges~\cite{Davies2016}.

Most spin wave generation techniques, 
except for the spin torque oscillator optimized by Bonetti et al.~\cite{Bonetti2009}, 
generate waves of frequencies below 20~GHz. 
On the other hand, acoustic wave generation by picosecond laser pulses 
has allowed for the investigation of waves 
in the gigahertz and even terahertz regimes~\cite{Matsuda2015, Huynh2008, Pu2005, Pezeril2016}. 
Resonant magnetization precession induced by ultra high frequency phonons has been reported recently~\cite{Akimov2015}. 
The interaction between laser-pumped acoustic waves and spin waves 
is under intense investigation now~\cite{Janusonis2016, Chang2016, Walowski2016, Linnik2011}. 

Originally considered by Kittel~\cite{Kittel1958}, the interaction of acoustic and spin waves  
has recently been modeled for parametric~\cite{Keshtgar2014, Zhang2016} and resonant~\cite{Dreher2012, Kamra2015, Gowtham2015} spin wave pumping. 
In the latter case coupling between the two types of waves is possible if their dispersion relations cross. However, such magnetoelastic waves exist only for a specific frequencies at crossings. Here we demonstrate, that it is possible to obtain magnetoelastic waves existing in broad range of frequencies in magphonic crystal and consider it as a way to generate high-frequency spin waves through acoustic waves.

In this paper we demonstrate theoretically that 
by optimizing the structural parameters of a 1D magphonic crystal 
it is possible to equalize the slopes of selected spin-wave and acoustic dispersion branches 
in the entire Brillouin zone. 
Then, the spin wave branch can be shifted to the frequency level of the acoustic branch 
by tuning the magnitude of the external magnetic field. 
This allows for strong magnetoelastic interaction 
in a broad range of frequencies and wave vectors. 
Using the plane wave method, we calculate the magnetoelastic dispersion relation 
and the relative amplitudes of magnetic and acoustic Bloch waves. 
Next, we perform finite-element time-domain simulations 
to demonstrate the application of this effect 
for efficient excitation of high-frequency short-wavelength spin waves 
by acoustic waves.

\begin{figure}
\includegraphics{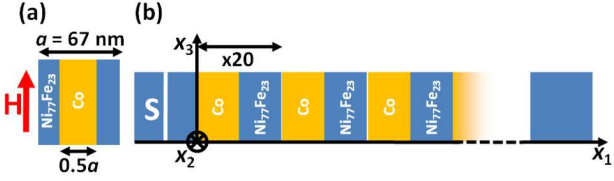}
 \caption{\label{Fig1} Geometry of the simulated system: 
 (a) unit cell used in PWM calculations; 
 (b) NiFe/Co magphonic structure considered in FEM simulations, 
 with a continuous acoustic wave excited by a point source~$S$ 
 and propagating along the $x_1$ axis. 
 Both acoustic and spin waves are damped near the boundaries of the system.}
\end{figure}

In our 1D model all the waves propagate along the $x_1$ axis 
in a ferromagnetic heterostructure composed of isotropic materials 
infinite in the $(x_2,x_3)$ plane (Fig.~\ref{Fig1}). 
The system is saturated magnetically along the $x_3$ axis by an external magnetic field~$H$. 
The evolution of the displacement component~$u_3$ 
and the dynamic magnetization components $m_1$ and $m_2$ 
in the exchange regime are described by the following equations~\cite{Comstock1963}:
\begin{equation}
\begin{split}
\dot{m}_1 = \omega_0 m_2 - \mu_0\gamma M_s \frac{\partial}{\partial x_1}l_{ex} \frac{\partial m_2}{\partial x_1}, \\
\dot{m}_2 = - \omega_0 m_1 + \mu_0\gamma M_s \frac{\partial}{\partial x_1}l_{ex} \frac{\partial m_1}{\partial x_1} - \gamma B \frac{\partial u_3}{\partial x_1}, \\
\rho \ddot{u}_3 = \frac{\partial}{\partial x_1} c \frac{\partial u_3}{\partial x_1}+\frac{\partial}{\partial x_1}\beta m_1,
\label{eq1}
\end{split}
\end{equation}
where $\omega_0=\gamma \mu_0 H$, $\gamma=176$ GHz/T is the gyromagnetic ratio, 
$M_s$ the saturation magnetization, $\mu_0$ the magnetic susceptibility of vacuum, 
$l_{ex}=2A/\mu_0 M_s^2$ the square of the exchange length, $A$ the exchange constant, 
$\rho$ the mass density, and $c\equiv c_{44}$ a component of the elastic tensor. 
Equations~(\ref{eq1}) are coupled by magnetoelastic terms 
with magnetoelastic coefficients $B$ and $\beta = B/M_s$. 

We used the plane wave method (PWM) for solving Eqs.~(\ref{eq1}) for an infinite 1D magphonic crystal (Fig.~1a)~\cite{Charles2006}. 
By using Bloch's theorem and expanding all the periodic functions into Fourier series 
we transformed the differential equations~(\ref{eq1}) 
into an infinite set of algebraic equations. 
In the computations we assumed $N=41$ reciprocal wave vectors ($N=41$ elements in the Fourier series) 
for a finite set of equations ensuring satisfactory convergence. 
The minimums of the matrix determinant of this homogeneous system of algebraic equations determine the dispersion relation. 

In the PWM calculations the amplitudes of $m_1$ and $m_2$ in Eqs.~(\ref{eq1}) 
were scaled by a factor $\kappa$ ($m_i = \kappa \tilde{m}_i$):
\begin{equation}
\kappa = \left(2\hat{A}k^2+\mu_0 H \hat{M}_s\right)^{-1/2},
\end{equation}
and the displacement $u_3$ by a factor $\kappa '$ ($u_3 = \kappa ' \tilde{u}_3$):
\begin{equation}
\kappa ' = \left(\hat{M}_s^2(\hat{\rho}\omega^2+\hat{c}k^2)\right)^{-1/2},
\end{equation}
where $k=q+G$ is the wave vector, 
$q$ a wave vector from the 1st Brillouin zone, 
and $G=2\pi n/a$ the reciprocal vector, with $n$ being an integer and $a$ the lattice constant. 
The hat symbols denote the weighted mean values in the effective homogeneous structure 
(here the weights are equal, since $ff=0.5$). 
These substitutions let us identify the spin or elastic character of the wave of frequency $\omega$  
for each wave vector~$k$ and each band in the dispersion relations presented below 
by the proportion of elastic energy carried by the wave: 
\begin{equation}
\frac{\hat{E}_{el}}{\hat{E}_m+\hat{E}_{el}} = \frac{\tilde{u}_3^2}{\tilde{m}_1^2+\tilde{m}_2^2+\tilde{u}_3^2}~,
\label{eq4}
\end{equation}
where $\hat{E}_{el}$ is the elastic wave energy density
\begin{equation}
E_{el} = \frac{1}{2}\left(ck^2+\rho\omega^2\right)u_3^2,
\label{eq2}
\end{equation} 
and $\hat{E}_m$ the spin wave energy 
\begin{equation}
E_{m} = \left(\frac{A}{M_s^2}k^2+\frac{\mu_0 H}{2M_s}\right)\left(m_1^2+m_2^2\right)
\label{eq3}
\end{equation}
in the effective homogeneous medium. 
The mode amplitudes $\left\{\tilde{m}_i, \tilde{u}_3\right\} = \sum_G \left\{\tilde{m}_i^G, \tilde{u}_3^G\right\}$ 
were found by the singular value decomposition routine from the Lapack library. 

We considered a magphonic crystal composed of 
permalloy (Ni\raisebox{-.4ex}{\scriptsize 77}Fe\raisebox{-.4ex}{\scriptsize 23}, NiFe) and cobalt 
with the filling fraction $ff=0.5$. 
Permalloy of this composition has a weak positive magnetostriction, 
while Co has a strong negative magnetostriction~\cite{Klokholm1981, Alberts1963}. 
In the calculations we assumed the following parameters of the NiFe constituent: 
$\rho = 8720$ kg/m$^3$, $c=50$ GPa, $A=13$ pJ/m, $M_s=760$ kA/m, $B=-0.9$ MJ/m$^3$. 
The Co parameters are $\rho = 8900$ kg/m$^3$, $c=80$ GPa, $A=20$ pJ/m, $M_s=1000$ kA/m, $B=10$ MJ/m$^3$.

The dispersion relations of spin and acoustic waves in homogeneous media (with magnetoelastic coupling neglected) 
are parabolic with a gap at the wave vector $k=0$ and linear without gap, respectively. 
By Bloch's theorem the introduction of periodicity into the structure 
will result in a periodic dispersion relation 
with a period equal to the reciprocal lattice vector, $f(q)=f(q+nG)$. 
This effect, known as the folding back to the first Brillouin zone, 
allows for multiple crossings of the spin-wave and acoustic branches, 
which can provide conditions favorable for magnetoelastic interaction between spin and acoustic waves. 
This interaction only occurs in close proximity to the crossing points, in narrow ranges of $f$ and $q$. 
The question arises whether the folding-back effect can be designed 
so that the spin-wave and acoustic dispersion relations overlap 
in a broad frequency and wave vector range. 
This can be realized if the dispersion curves
of acoustic and spin waves are almost parallel in some range of~$k$. 

The values of the lattice constant for which the spin-wave branch is parallel to the acoustic branch 
in a magphonic crystal can be estimated by finding pairs $(k,f)$ in the dispersion relations of spin and acoustic waves  
corresponding to equal group velocities in the effective homogeneous medium. 
We select a matching point $k$ in the middle of the Brillouin zone,~$n\pi/2a$. 
The resulting condition for a homogeneous material with the effective parameters of the NiFe/Co magphonic crystal is:
\begin{equation}
a_n=(2n+1)\frac{\pi \hat{D}}{\hat{v}},
\end{equation}
where $D=2A\gamma /M_s$ and $v=\sqrt{c/\rho}$ is the acoustic wave velocity. 

The calculated lattice constant values~$a_n$ 
corresponding to equal group velocities of both excitations 
are $7.6$~nm, $22.9$~nm, $38.1$~nm and $53.3$~nm for $n=0,1,2$ and 3, respectively. 
However, the spin-wave frequencies lie above the acoustic frequencies 
and therefore cannot be aligned with the latter by external magnetic field. 
The first spin-wave branch parallel to an acoustic branch and lying below it appears for $a_4=68.6$~nm. 
This lattice constant value was manually optimized to $67$~nm. 
Under these conditions the fifth spin-wave branch overlaps with the third acoustic branch 
in a $100$~kA/m magnetic field. 

\begin{figure}
\includegraphics{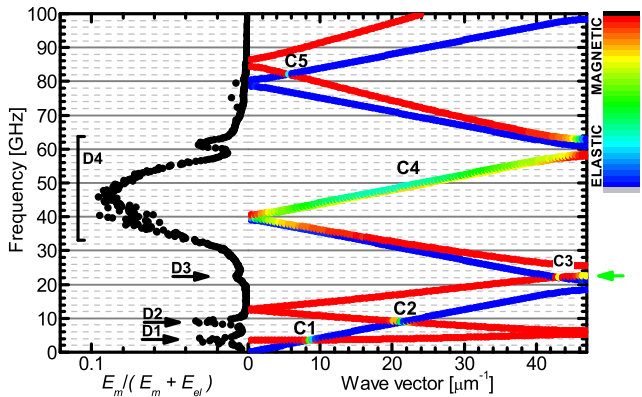}
 \caption{\label{Fig2} (right) Dispersion relation of the NiFe/Co magphonic crystal calculated by the PWM. 
 In the color scale, blue and red correspond to acoustic and spin waves, respectively. 
 Green indicates coupled magnetoelastic waves. 
 (left) Proportion of magnetic energy accumulated in the system in 1~ns of excitation by acoustic wave of a given frequency.}
\end{figure}

Indeed, also in the numerical results overlapping is observed with these parameters. 
In Fig.~\ref{Fig2}, on the right, we present the dispersion relation of the considered periodic system calculated by the PWM. 
Acoustic and spin-wave branches cross at four frequencies: 
3.6~GHz, 9~GHz, 22.1~GHz and 82.1~GHz, labeled as C1, C2, C3 and C5, respectively. 
The third acoustic branch overlaps with the fifth spin-wave branch 
throughout the Brillouin zone in the frequency range from ca.~40~GHz to 60~GHz, labeled as C4. 
The color of the dispersion points, 
indicating the proportion of elastic and magnetic
energy of the wave, described by Eq.~(\ref{eq4}), 
shows that the modes of these frequencies are hybridized magnetoelastic waves.

\begin{figure}
\includegraphics{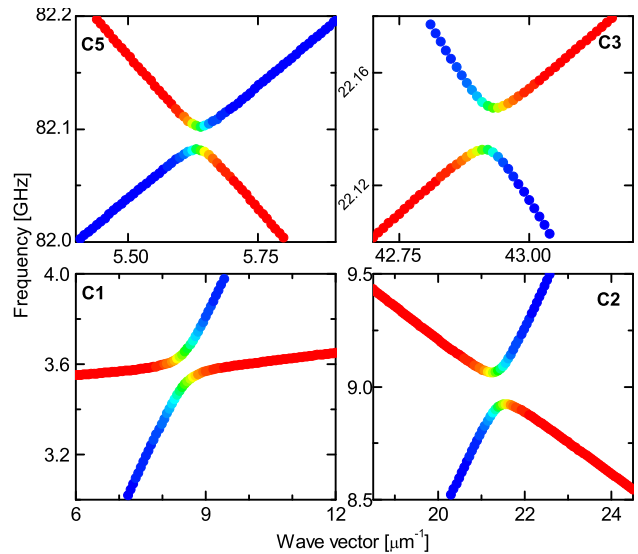}
 \caption{\label{Fig3} Zoomed anti-crossings C1, C2, C3 and C5 (indicated in the dispersion relation in Fig.~2) 
 in the NiFe/Co magphonic crystal.}
\end{figure}

Figure~\ref{Fig3} presents zoomed parts of the dispersion plotted in Fig.~\ref{Fig2} 
around crossings C1, C2, C3 and C5. Clearly, the spin-wave and acoustic dispersions anti-cross at these points. 
Measured as the frequency difference between branches 
at the point of mixed polarization (50\% of elastic and 50\% of magnetic energy),
the band gap widths at anti-crossings C1, C2, C3 and C5
are 130~MHz, 156~MHz, 16~MHz and 21~MHz, respectively.  
At C4 (in the range from 42~GHz to 56~GHz) the branches are ca.~400~MHz apart, 
while in the absence of magnetoelastic coupling they overlap almost completely.

To test the possibility of excitation of spin waves via acoustic waves in a 1D magphonic crystal 
we performed time-domain simulations of a finite structure equivalent to that considered above. 
We used the COMSOL Multiphysics environment for solving the system of equations~(\ref{eq1}) by the finite element method~(FEM). 
The considered geometry is shown in Fig.~\ref{Fig1}b. 
The magphonic crystal comprises 20 repetitions of the NiFe/Co unit cell, bounded by homogeneous NiFe. 
Areas of increased damping are introduced at the ends of the structure 
to suppress wave reflections from the edges. 
A continuous acoustic vibration was excited for 1~ns with a source, indicated in Fig.~\ref{Fig1}, 
placed in homogeneous permalloy at a distance of 250~nm from the magphonic crystal. 
We determined the ratio~$E_m/(E_m+E_{el})$ of the accumulated magnetic energy 
to the total energy in the whole simulated structure (Eqs.~(\ref{eq2}) and (\ref{eq3})). 
Independent calculations were performed for each excitation frequency. 

In Fig.~\ref{Fig2}, on the left, the proportion of magnetic energy accumulated in the system, 
obtained from the time-domain simulations, is plotted versus the frequency of the acoustic wave source. 
These results can be directly compared with the PWM dispersion relation. 
Three maximums of the magnetic energy, labeled as D1, D2 and D3, occur at 3.6~GHz, 9~GHz and 22.6~GHz, respectively.  
Moreover, a broad D4 peak occurs at 46~GHz, extending slightly above the band gap at 59~GHz 
and below the narrow band gap at 40 GHz. 
By comparing these frequencies with those of the anti-crossings shown on the right in Fig.~2 
we find that the D1 and D2 maximums correspond to anti-crossings C1 and C2. 
The broad D4 peak corresponds to the overlapping of spin-wave and acoustic branches at C4. 
The D3 maximum, however, seems to be an effect of the coupling 
between branches at the Brillouin zone boundary rather than the C3 crossing, 
indicated by a green arrow in Fig.~\ref{Fig2} 
(see below for further discussion).

The maximal proportions of magnetic energy stored in the system at D1, D2, D3 and D4
are 0.031, 0.034, 0.008 and 0.1, respectively.  
When compared with these proportions, the band gap widths discussed above 
appear as a measure of magnetoelastic coupling. 
For example, the magnetic energy at D4 is almost three times higher than at D1; 
also the gap between branches at C4 is about three times larger than that at the C1 anti-crossing. 
On the other hand, the band gaps at the C3 and C4 anti-crossings are an order of magnitude smaller 
than those at the C1 and C2 anti-crossings. 
This is why they are not observed in the spectra in Fig.~\ref{Fig2}, left. 
The 22.6~GHz maximum of 0.008 results rather from the branch coupling near the Brillouin zone boundary. 

\begin{figure}
\includegraphics{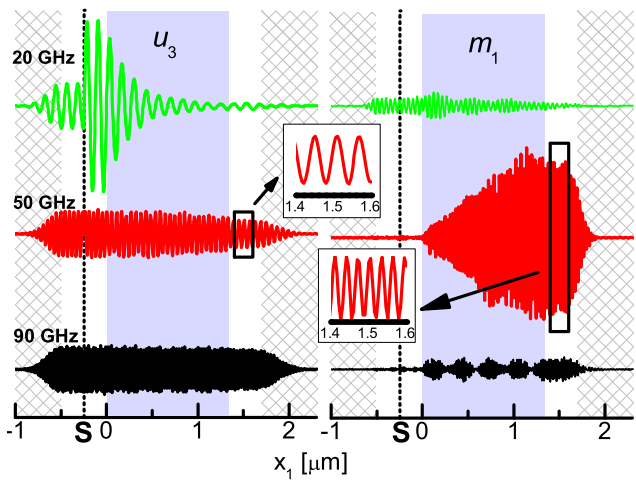}
 \caption{\label{Fig4} Acoustic (left) and spin-wave (right) signals 
 after 1~ns of excitation by an acoustic wave from a source point~$S$
 at 20~GHz, 50~GHz and 90~GHz. 
 Insets show enlarged plots of the 50~GHz acoustic and spin-wave signals beyond the magphonic crystal. 
 Uniform gray and patterned areas represent the magphonic crystal and damping edges, respectively. 
 The amplitude ratio of the acoustic and spin waves is preserved.}
\end{figure}

It is worthy of notice that in a magphonic crystal 
overlapping of spin-wave and acoustic dispersion curves 
is possible under much lower magnetic fields than in a homogeneous material. 
For example, in homogeneous YIG the spin-wave and acoustic dispersion curves touch  
under a 2.5~T field~\cite{Kikkawa2016}, 
which is one order of magnitude higher than the field necessary in the case of the magphonic crystal proposed above.

Figure~\ref{Fig4} shows acoustic and spin waves 
after 1~ns of excitation by an acoustic source at three different frequencies. 
At 20~GHz the acoustic wave is reflected by the magphonic crystal, 
since the band gap at this frequency prohibits the propagation of acoustic waves 
(see Fig.~\ref{Fig2}, right); consequently, the generated spin wave has a low amplitude. 
At 50~GHz, a frequency in the broad band of magnetoelastic interaction C4,   
the amplitude of the acoustic wave decreases as the wave propagates through the magphonic crystal, 
while the amplitude of the spin wave is significantly amplified. 
At 90~GHz the spin wave generation is suppressed again, because of the low magnetoelastic interaction at this frequency.

The insets in Fig.~\ref{Fig4} present enlarged plots of the 50~GHz acoustic and spin waves  
that have left the magphonic crystal. 
The wavelength of the spin wave is sensibly shorter than that of the acoustic wave. 
The precise values are 60~nm and 30~nm for the acoustic and spin waves, respectively. 
This is due to the fact that at C4 
a spin wave branch from the fifth band overlaps with an acoustic branch from the third band of the magphonic crystal. 
Generally, in the C4 band the wavelengths of spin waves are in the range from 27~nm to 33~nm, 
while those of exciting acoustic waves range from 45~nm to 67~nm. 

In summary, we have demonstrated that the periodicity of a magphonic crystal 
allows for coupling between acoustic and spin wave dispersion branches in the whole Brillouin zone. 
Therefore, magphonic crystals with such a property are promising candidates for spin wave generation. 
They can be used for efficient generation of spin waves by acoustic waves 
in a broad band of frequencies and wave vectors and with reasonable magnitudes of the external magnetic field. 
The generated waves can have a nanometer wavelength much shorter than the source wave. 

The efficiency of spin wave generation in practical applications
remains an open question in the context of optimization of the magphonic crystal size versus wave attenuation. 
We used spin waves in the high-frequency exchange regime, which is desirable, but hard to attain experimentally. 
By contrast, acoustic waves with frequencies of that order can be generated by laser pumping. 
On the other hand, it seems possible to achieve similar branch overlapping with low-frequency acoustic bands, 
where the magnetostatic interaction contributes to the spin wave dynamics. 
However, testing this idea requires more complex numerical simulations. 
Of crucial importance is also the investigation, in the context of practical applications,  
of higher-dimensional structures, in which, e.g., surface waves can be considered.

\begin{acknowledgments}
We would like to thank Pawel Gruszecki for valuable comments during the preparation of this paper.
The study has received financial support from the National Science Centre of Poland 
under Sonata BIS grant UMO-2012/07/E/ST3/00538 
and the EU’s Horizon 2020 research and innovation programme under Marie Sklodowska-Curie GA No.~644348 (MagIC).
\end{acknowledgments}

\end{document}